\documentstyle[11pt,a4,times,amssymb]{article}

\newcommand{\CP} {{\mathbb C\mathbf P}}
\newcommand{\Z} {{\mathbb Z}}

\begin{document}

\begin{titlepage}
\null \vskip -0.6cm
\hfill PAR--LPTHE 99--25

\hfill chao-dyn/9912012

\vskip 1.4truecm
\begin{center}
\obeylines

        {\Large	Algebraic Entropy of Birational Maps
                with Invariant Curves}
\vskip 1cm
M.P. Bellon
{\em Laboratoire de Physique Th\'eorique et Hautes Energies
Unit\'e Mixte de Recherche 7589, CNRS et Universit\'es Paris VI et Paris VII.}
\end{center}

\noindent{\bf Abstract}: 
The relevance of the algebraic entropy in the study of birational
discrete time dynamical systems highlights the need to relate it to other
characteristics of these systems.
In this letter, two complementary proofs are given that
the foliation of the space by invariant curves
implies that the algebraic entropy is zero.  

\vfill

\begin{center}

\hrule \medskip
\obeylines
Postal address: %
Laboratoire de Physique Th\'eorique et des Hautes Energies
Universit\'e Pierre et Marie Curie, bo\^\i te postale 126.
4 Place Jussieu/ F--75252 PARIS Cedex 05

\end{center}
\end{titlepage}

\section{Introduction}

Dynamical systems come with a variety of behaviours, which have been the
object of many studies.  In particular, the long term predictability
is a subject of interest since the first considerations
on the solar system and received their modern formulation from Poincar\'e
a century ago~\cite{Po92}.  

It has been shown
\cite{BeMaVi91b} that birational maps have a great variety of dynamical 
behaviours, in spite of their apparent simplicity:  We need only the
basic four operations to compute the successive points and we can even
dispense with the division if we keep with homogeneous coordinates.
The purpose of this note is to relate different characterizations
of dynamical systems for birational maps.

The first one is the existence of invariants,  which allow for a Liouville
like form of integrability.  An other one is the sensitivity to initial
conditions, which is related to the derivative.  The last one is the
``algebraic entropy''~\cite{BeVi99}.  Common factors in the
homogeneous coordinates of the image of a point result in simplifications
of its expression and decrease the degree of the $n$-th iterate 
of the transformation from the $d^n$ value obtained without simplifications.
This notion is directly linked with Arnold's complexity~\cite{Ar90},
since the degree of a map gives the number of intersections of
the image of a line and a hyperplane.

A zero value of the algebraic entropy corresponds to a sub-exponential
growth of the successive degrees and in all known cases to a degree 
depending polynomially on $n$.  A first proof of such a polynomial bound on the
degrees was given in~\cite{BoMaRo94}.  It has been observed on numerous
examples through the determination of the images of a generic line.  This
would result from the conjectured
existence of recurrence relations on the degrees. 

This work is motivated by the observation that whenever the iterates
are confined to a low dimensional subvarieties by conserved quantities,
it has been possible to find that the algebraic entropy is zero.
Moreover, in many cases the growth is simply quadratic.

Here I show how this polynomial behaviour of the degrees can 
be derived from the existence of invariant subvarieties which are 
abelian varieties.  In the simplest case of rational or elliptic 
curves, the exact exponent will be proven.

The first section will precise what is called integrability in this
context of discrete time dynamics.  Next will come an analytic derivation
of a quite general bound on the growth of the degrees of the iterates,
based on the expression of the degree of a variety from its volume.
The last section makes use of the notion of addition on an elliptic curve
to build iterates with degrees of quadratic polynomial growth.
The practical use of such constructions and the foreseeable extensions are
the subject of the conclusion.

\section{Action-angle variables}
Even if it is intuitively clear that integrability has to do with the
existence of a sufficient number of conserved quantities, the question
remains of what is sufficient.  A discrete time analogue of Hamiltonian
mechanics has been proposed~\cite{MoVe91}, allowing for a direct
generalizations of the answer provided by Liouville~\cite{Li55}.  However,
the resulting class of systems is too restrictive for our porpoise and we
cannot rely on a definitive answer. 

Since even two-dimensional algebraic varieties can have birational maps
with complex dynamical behaviour, the only safe case is the one 
of one-dimensional invariant varieties.  In this case, the induced
birational map on the invariant varieties are in fact holomorphic:
Singular varieties of birational maps
are of codimension at least two and there are no such subvarieties in curves.
It remains to classify
automorphisms of infinite order of the curves, which exist only for curves
of genus 0 or 1.  Higher genus curves have but a finite number of
automorphisms.  The only possible automorphisms have therefore finite order
and as dynamical systems, they are periodic.

In the genus 0 case, the curve is the Riemann sphere and the automorphisms
are homographic transformations.  By a reparameterization, we can always
obtain one of the cases $g(z) = z + \alpha$ or $g(z) = \lambda z$,
according to the number of fixed points of the transformation.  In the
multiplicative case, we can describe it as a translation if we set $z =
\exp t$.  The point depends on the parameter $t$ with the period $2\pi I$
and now the transformation is a simple addition on $t$.

Non-singular genus 1 curves are holomorphically isomorphic to a torus
$\mathbb C/\mathbb Z + \tau \mathbb Z$.
There is therefore a parameterization of these curves by doubly periodic
functions of $z \in \mathbb C$.  In terms of this parameter, the infinite
order automorphisms are translations $z \to z+\beta$.

In all these cases, the map is described by a simple translation in a
parameter which can be one to one, simply periodic or doubly periodic.
The $n$-th iterate of the map can be simply expressed as the translation by
the same parameter multiplied by $n$.  However there is a price to pay,
especially in the genus 1 case, since we have no longer algebraic
functions, but a parameterization by Weierstra\ss\ functions or other elliptic
functions.

Nevertheless, the scheme is that in adapted coordinates, the
transformation takes the simple form:
\begin{equation}\label{phi}
        \phi(X, \theta ) = (X,\theta + \sigma(X)),
\end{equation}
with $X$ designing the invariants of the integral curve and $\theta$ a
parameterization of this curve.
In these coordinates, the $n$-th iterates of $\phi$ is easy to write:
\begin{equation} \label{phin}
        \phi^n(X, \theta ) = ( X, \theta + n \sigma(X) ).
\end{equation}
From the explicit form~(\ref{phin}) of~$\phi^n$, its differential is
deduced to be:
\begin{equation}\label{dphin}
        d ( \phi^n ) = \pmatrix{ 1&0\cr n\sigma'&1\cr}.
\end{equation}
The important property is that the matrix elements grow at most linearly in
$n$, giving a computable behaviour for long times, since numerical errors
get multiplied by small numbers.  In generic systems, there would be
hyperbolic fixed points of $\phi^N$.  At these points, the differential of
$\psi^{kN}$ grows exponentially.  Small errors in
the first few iterations can have sizable effects after a moderate number of
iterations.

There are natural generalizations of this scheme for higher dimensional
varieties, but in the absence of an equivalent to the symplectic structure
which is essential in the theorem of Liouville, they will not be necessary
outcomes of the existence of invariant varieties.  Even in a two-dimensional
variety with rational parameterization, the restriction of a birational
map can have complex behaviours.  And it is not even sufficient that the map 
restricted to the invariant varieties has no singularities, since a torus of two 
complex dimensions can have exponentially diverging trajectories.  
An example is the following bijective
transformation of ${\mathbb C}^2/ \Z^2 + \tau \Z^2$ (the
product of two elliptic surfaces with the same modulus):
\begin{equation} \label{hyper}
(\theta_1,\theta_2)\rightarrow 
(2 \theta_1 + \theta_2, \theta_1+\theta_2).
\end{equation}

The following section is equally valid for the case of abelian
invariant varieties if the restriction of the map to these varieties is a
simple translation.  The formulas~(\ref{phi},\ref{phin},\ref{dphin})
apply with a multidimensional $\theta$.
Examples of such systems were shown to exist in~\cite{Ko93}, with a map
which reduces to an addition on the Jacobian of a curve.  It is however not
clear if these systems can be expressed as maps on projective spaces,
since they are {\em a priori} defined on the quotient of a linear space of
matrices by gauge transformations.  In the case which gives one-dimensional
Jacobians, it is possible to fix the gauge and the system reduces to
the one associated to the height-vertex model of Baxter~\cite{BeMaVi92}.
In the other cases, it is not clear whether such a choice is possible in
general, without restricting to a lower dimensional part of the Jacobian.
Part of the symmetry group of a three dimensional model of statistical
mechanics gives a dynamical system in $\CP^9$ with two-dimensional
invariant varieties~\cite{BeBoMaVi93}.  From the images of the orbits, it
seems that the restriction of the map on an invariant variety is a shift on
a torus of complex dimension two, but this has yet to be proved.

\section{Analytic method.}
\subsection{Fixed points}
An immediate consequence of the form~(\ref{phin}) of the compositions of
$\phi$ is that the fixed points of the iterates are not isolated, but form
whole varieties.  If $p$ is a fixed point of $\phi^n$, it means that
$n\sigma(X)$ is one of the periods of the parameterization of the invariant
curve it belongs to and all points of this curve are fixed points of $\phi^n$.
Moreover, the equation determining the fixed point variety reduces to 
$\sigma_n(X) = 0$.  This is as many equations as there are components in
$\theta$.  The codimension of the fixed point varieties is therefore the 
dimension of the invariant varieties.  

In the bidimensional case, the fixed point variety is of the same
dimension as the invariant varieties and reduces to a product of
invariant varieties.  This factorization is however not straightforward
since the corresponding values of the invariants are generically
irrational algebraic numbers.   For the determination of the individual
invariant varieties, the factorization of the equation of the fixed
point variety must be done in an algebraic extension of $\mathbb Q$
which has to be determined.

In higher dimensions, the fixed point variety would correspond to a
variety in the space of invariants, so that it would be still more
difficult to deduce the equations of individual invariant varieties.
However, the main problem in determining invariant varieties is to find
their covariance factor, that is the factor they get multiplied by when
applying the transformation.  The fixed point varieties are invariant
and their covariance factors are powers of the covariance factors of the
invariant varieties.

Moreover, the study of the differential of the map at fixed points should 
be a good starting point for a proof of the existence of invariant varieties.

\subsection{Analytic proof of a polynomial growth of the
degrees}\label{anal}
The degree $d$ of an algebraic subvariety is proportional to its volume in the
projectively invariant K\"ahler metric of $\CP^N$~\cite{Mu95}.  The basic
idea is that the volume forms induced on complex subvarieties by a K\"ahler
metric are simply exterior powers of the associated K\"ahler form.  The
K\"ahler form being closed, this volume is invariant by deformations and in
particular, can be computed from the retraction of the variety to a linear
space $L$ of the same dimension.  In a generic situation, points outside a
lower dimensional subspace of $L$ have $d$ preimages by the retraction,
so the volume of a variety is $d$ time the volume of a linear space.
Since all linear spaces are related by $U(N+1)$ transformations which leave
the metric invariant, the volume of a linear space is a universal constant.

This gives an analytic way of computing the degrees.  The degree of
the map $\phi$ is the degree of the image of a generic complex line $L$ and
this degree can be computed by an integration, since it is proportional to
the area $S$:
\begin{equation}\label{Vol}
         S = \int_{\phi^n(L)} \omega = \int_L \phi^n_\star \,\omega.
\end{equation}
In this formula  $\omega$ denotes the K\"ahler form.  The definition of
its pull-back $\phi^n_\star\omega$ involves the differential $d \phi^n$ of
$\phi^n$ to
transform the tangent vectors of $L$ into tangent vectors of its image.
More precisely, the form $\omega$ and the differential $d\phi$ are sections
of fiber bundles with values at a point $x$ denoted respectively by
$\omega_x$ and $d\phi_x$.  $\omega_x$ is a bilinear form on the tangent space
$T_x M$, $d\phi_a$ is a linear map from $T_a M$ to $T_{\phi(a)}M$.
$\phi_* \omega$ is given by
\begin{equation}
        (\phi_*\omega)_a(X,Y) = \omega_{\phi(a)} (d\phi_a X, d\phi_a Y),
\end{equation}
with $X$, $Y$ in $T_aM$.
From the formula~(\ref{dphin}) for $d\phi^n$ follows that we have a simple
dependence on $n$ of the integral in~(\ref{Vol}).  The K\"ahler
form must be evaluated in the variable point $\phi^n(a)$, with the
tangent vector expressed on a basis associated to the action-angle
coordinates.   $\CP^N$ is compact, $\omega$ is bounded in any affine
coordinate patch from its explicit expression and the change from
action-angle variables to projective coordinates is smooth, so that 
$\omega$ remains bounded when expressed on an adapted basis of the tangent
space.
The boundedness of $\omega_x$ 
is then sufficient to conclude that the degree of $\phi^n$ is less
than some constant times $n^2$.  

\section{Algebraic proof}

\label{alg}
\subsection{Generalities}
In the preceding section, I showed limits on the degrees of the iterated
maps.  The proof however does not show how to build transformations 
of the given degree.  This will be remedied in this section.  

The basic idea is that eq.~(\ref{phin}) can be expressed algebraicly, without
the introduction of a specific parameterization of the invariant variety.
Since every analytic operation on an algebraic variety is algebraic, 
$A + B$ is an algebraic function of 
$A$, $B$ and parameters describing the invariant variety.  The algebraic
translation of eq.~(\ref{phin}) is made of operations independent of $n$
and the computation of $n\sigma$.  The whole $n$ dependence comes from this
operation and we want to find a good bound of the degree of this operation.
All other operations can contribute an overall factor or some additional
constants but cannot modify the growth behaviour.  Computing $2k\, X$ from
$k\, X$ is a simple addition.  $2^p \, X$ can therefore be computed by $p$
additions and for a general $n$, the number of additions to perform to
calculate $n\,X$ is of order $\log_2 n$.
If the degree of the result is multiplied by a constant $r$ by each
addition, the degree of $n\,\sigma$ will be of order $r^{\log_2 n}$, which
is equal to $n^{\log_2 r}$.  We therefore have built an expression for the
image of a point with a degree satisfying the required bound.

In fact, this discussion is somehow naive.  The invariant variety
depends on the starting point and care must be taken of the dependence of the
addition on the parameters of the invariant variety, which will in turn be some
polynomial expressions of the coordinates of the starting point.  
Going from the point $pz$ to the point $2pz$ is an
operation of degree $d$ in the coordinates of the point $pz$ and of some
degree $l$ in the coordinates of the initial point.  Let us call $m_k$ the
degree of the operation $z \to 2^k z$.  $m_k$ satisfies the following
relation:
\begin{equation}
 m_{k+1} = d m_k + l.
\end{equation}
It is elementary to verify that $m_k$ is also bounded by $C d^k$, except in
the trivial case $d=1$.

The degree of the iterate of order $n$ of an integrable mapping
is therefore bounded by a polynomial function of $n$.
In the following, the precise form of this bound will be established in the
case of curves.  A generalization for higher dimensional cases is in
principle straightforward, but rather involved.

\subsection{Curves}
The case of rational invariant curves can be easily settled, 
since the parameterizations of the invariant curves
are rational.  A birational change of variable allows for a product structure
$\CP_1$ times the space of the invariants.

In the case where the $\theta$-variable is directly the rational parameter,
the expression of $\phi^n$ can be directly read from~(\ref{phin}). In this
case, the successive iterates have all the same degree.

In the other case, the transformation takes the form $t\rightarrow \lambda(X) t$.
The iterates depend on $\lambda(X)^n$ and the degree of the transformation
is linear in $n$.  

In the case of elliptic invariant curves, the choice of a special point $O$
allows for the identification of the curve with its Jacobian.
This determines the addition on the curve: the sum $S$
of the points $A$ and $B$ is the
second zero of a meromorphic function with simple poles at these two points
and a zero at the origin $O$.  Such a meromorphic function can be obtained 
as the ratio of sections of a line bundle.   In the simplest cases, these sections 
are simply linear functions of the homogeneous coordinates.  Their zeroes are
given by the intersection of the curve and some hyperplane.  

There are two simple algebraic descriptions of an elliptic curve.  The
first is a degree 3
plane curve, the second is the intersection of two quadrics in
three-dimensional space.  In both cases, the explicit determination of
multiples of a point will allow for a quadratic bound on the degrees of the
successive iterates.

Otherwise, we could parameterize the curves by doubly periodic
functions, Weierstra\ss\ function for the degree 3 plane curve or
$\theta$-functions for the biquadratic.  Then addition formulas for these
functions would translate in algebraic formulas for the determination of
the sums, but I do not want to introduce any considerations on 
these special functions.

Higher degree representations are singular with multiple points.   
These multiple points correspond to multiple values of the parameter
of the curve and are sent to a number of differing points.  They are 
therefore singular points of the transformation and it is possible to make
birational transformations which resolve those multiple points without
introducing other multiple points.

\subsection{Degree 3 plane curve}
Let us consider a curve with the equation:
\begin{equation}\label{deg3}
	y^2 = P_3(x).
\end{equation}
$P_3$ is a degree three polynomial with no multiple roots.  The natural choice 
for the origin $O$ in this case is the point at infinity.  This point is common
to all invariant curves so that the parameter $\sigma(X)$ cannot be obtained
directly as the image of the origin $O$, but must be computed as $\phi(P) -P$
for a generic point $P$ on the curve.

Affine functions of $x$ and $y$ are sections of a line bundle when
restricted to the curve.  The zeros of these sections are given by the
intersection points of a line and the elliptic curve: there are three of
them.  

To build a meromorphic function with poles in $A$ and $B$, we first
consider the line through $A$ and $B$: it will cross the curve in a third
point $M$.  We will take now a line going through $M$ and the origin $O$: I
claim that the third intersection point of this line with the curve is the
sum $S$.  Indeed, taking the ratio of the two above mentioned linear
functions, we get a meromorphic function on the curve with poles in $A$ and
$B$ and zeros in $O$ and $S$.  The zeros in $M$ of the two linear functions
cancel each other.  Lines going through $O$
have equation $x=x_0$ and intersect the curve in two points of the form
$(x,\pm y)$.

When defining the double of a point, we need to parameterize the tangent to the curve
at the given point.  It will be given by $(x+2\alpha y, y + \alpha P'_3(x))$.
Assuming a reduced form for $P_3$, that is $P_3=x^3+ax+b$, the equation for
$\alpha$ to get a point on the curve reduces to:
\begin{equation}
        P'_3(x)^2 \alpha^2 = 8\alpha^3y^3 + 12 \alpha^2 x y^2.  
\end{equation}
The non zero solution for $\alpha$ can be substituted and $y^2$ expressed
in terms of $x$ from the equation (\ref{deg3}).  Remembering that 
the point $2z$ has the opposite $y$ from the intersection, we obtain the
following coordinates for it:
\begin{equation}
\label{double}
        \left(-2x+{P'_3(x)^2\over 4 P_3(x)},
        -y \Big(1-{3xP'_3(x)\over 2 P_3(x)}
        +{P'_3(x)^3\over 8 P_3(x)^2}\Big) \right).
\end{equation}
The important point is here that the new $x$ is of degree 4 in the old $x$,
without any dependence on $y$.  As in the preceding section, it is possible
to deduce that the degree of $x$ for the $k$-th iterates is at most
quadratic in $k$.  The degree of $y$ does not seem to be as easily bounded.
But the very expression of the new $y$ shows
that its degree is the degree of the old $y$ plus 6 times the degree of the 
old $x$.  The degree of the variable $y$ after a number of steps will be 1
plus some times the sum of the degrees of the preceding $x$.  This is the
sum of a geometric series, so that it will also be of order $4^p$, apart
from some constant prefactor.

In this case, the integer $d$ is 4, that is to say the bound
derived in the preceding paragraph is a quadratic one.  

\subsection{Biquadratics in 3 dimensional space.}

The other possible equation for a non-singular elliptic curve is that
of a biquadratic in a three dimensional space.  By a proper choice of
coordinates and of combination of the equations, they can be brought to the
form:
\begin{eqnarray}
        x^2 - a y^2 -b z^2 &=& 0, \nonumber \\
        t^2 - b y^2 -a z^2 &=& 0.
\end{eqnarray}
This readily shows the symmetries of the curve, since these two equations
are invariant by the change of sign of any of the coordinates.
Since the change of the sign of all coordinates is the identity in 
projective space, this gives a group $\Z_2 \times \Z_2 \times \Z_2$ of
symmetries.  Section of the tautological bundle are just given by a linear
form in $x,y,z,t$.  They are defined up to a factor by a plane in
this three-dimensional space, that is by three points.  Intersection of a
plane with the curve will generically consist in 4 points.  Computation in
the Jacobian can easily be done if we take as zero a point with one of the
coordinates 0.  There are four of these points on any curve, but they are 
related by the afore mentioned symmetries.  

A plane is defined by three points, so that we choose a point $R$ on the
curve which will belong to the planes corresponding to the two factors
defining a meromorphic function.  The plane defined by $A$, $B$ and
$R$ will have a fourth intersection with the curve, $H$, and the plane 
going through $H$, $R$ and $O$ will meet the curve at the sum $S$:
when taking the quotient, the zeroes in $H$ and $R$ cancel and we get
the required meromorphic function with the divisor $A+B-S-O$.

$R$ should be chosen to be one of the points with the same zero as the
origin, since in this case $S$ is simply related to $H$.  In this case, the
direction of the line $OR$ is one of the coordinate axis.  It is either the 
axis where the coordinates of $O$ and $R$ differ by a sign or the axis
of the null coordinate in the case $R$ and $O$ coincide, since the
tangent in this point is simply giving by the vector whose
only non-zero component is on the null coordinate.  Then if we take for
$S$ the point obtained from $H$ by negating the coordinate along the
axis $OR$, $HS$ and $OR$ have the same direction and they are coplanar
as two parallel lines define a plane.  A remark which be useful in the 
sequel is that if $O$ is the point $(x_0,y_0,z_0,0)$ adding the point
$O_z=(x_0,y_0,-z_0,0)$ to a point is simply changing the sign of the $z$ and
$t$ coordinates.  With obvious notation, the addition of the $O_x$ and $O_y$
points are similarly obtained as symmetries of the curve changing the sign
of two of the coordinates.  Since the negative for the addition on the
curves is obtained by changing the sign of the $t$ coordinate, all
obvious symmetries of the curve can be simply obtained by a combination
of the addition of one of the points $O_i$ and the opposite.

The obvious way of adding multiple times the same point is to 
define a plane which is tangent to the curve at this point, but in
this case, we do not end with a result of sufficiently low degree.
It is however possible to consider the plane going through 
the original point and two of its transforms by symmetries to compute a triple.
The plane going through $A$, $A+O_x$ and $A+O_y$ intersects the
curve in the point $-3 A + O_z$ ($O_x+O_y=O_z$).

If we start from three points $A_1$, $A_2$ and $A_3$ on the curve, the plane
they define is parameterized by the linear combination
$ \alpha_1 A_1 + \alpha_2 A_2 +\alpha_3 A_3 $.  When substituting in the
equations of the curve, the $\alpha_i^2$ terms disappear since the $A_i$'s
satisfy these equations and we have two linear equations for the three
products $\alpha_1 \alpha_2$, $\alpha_1 \alpha_3$ and $\alpha_2 \alpha_3$,
whose coefficients are bilinear in the coordinates of the points.
These products are then of degree 4 in the points, the $\alpha_i$ of
degree 8 and the final point is of degree 9.  We obtain a variant
of the result we announced.  Here, we proceed by tripling and the operation
is of degree 9, which is the cube of 3, so that we still obtain a
quadratic bound for the degrees of the iterations of the maps.  

There still remains the problem that in fact, the origin $0=(x_0,y_0,z_0,0)$ 
of the curve is not algebraicly determined.  
$x_0^2$, $y_0^2$ and $z_0^2$ are expressed as quadratic polynomials
of the initial point, but the actual coordinates are square roots.  
We will have to keep track of the usage of each of
these quantities to be sure that the end result only depends on even powers
of these coordinates, so that the result is truly an algebraic one.  Here
again, the symmetries of the curve give us the possibility to simply derive
what happens if we change the sign of one of these coordinates.  Changing
the sign of $z_0$ for example, is changing $O$ to $O_z$.  Multiplying the
image of the origin will give the iterated image of this point.  Then to
obtain the image of the starting point $P$, we have to add $P$ with the
image of the origin.  But this addition depends on the origin, since
$A+B$ depends on the fact that $A+B-S-O$ is a principal divisor.  The
change in the origin completely compensate for the change of the point.
The conclusion is that the calculation does not depend on the possible
choices of signs.  Except from a global sign factor, the calculated point
is then invariant, so that it only depends on even powers of the 
coordinates of the origin and is therefore an algebraic function of the
starting point.  
This is an
elementary case of Galois theory, which says that element of an algebraic
extension which are invariant by all elements of the automorphism group
(the Galois group) are elements of the base field.

\section{Conclusion}
Integrable mappings have been proven to have zero algebraic entropy. 
Moreover, the observed quadratic polynomial growth of the degree
is seen to be generic from the results of sec.~(\ref{anal}).

This step forward in the study of the dynamical behaviour of birational
maps urges to find a reciprocal result.  However the case of the discrete
Painlev\'e system studied in~\cite{BeVi99} shows that a zero algebraic
entropy does not imply the existence of invariants, even in a
two-dimensional setting.  Maybe this is due to the non-autonomous character
of this map, which implies that fixed points of $\phi^{k}$ are not fixed
points of all the transformations $\phi^{kp}$.

The derivation of the higher order iterates sketched in
sec.~(\ref{alg}) has been useful for an alternate proof of the polynomial
growth of the degrees.  We can wonder whether this calculation can be put
to more practical uses.  The logarithmic growth of the number of operations
should in any case become a clear advantage both for the precision of the
calculation and the required time.  

The usefulness of the fixed point varieties for a characterization of the
invariant varieties should be probed.  But for higher dimensional systems,
the explicit equations are very large and difficult to obtain with current
algebraic manipulation programs.  It should be of interest to obtain
such varieties from implicit equations, i.e., without trying to have a
formula for the $\phi^n$.  Otherwise, intersection of some
low dimensional variety with the fixed point varieties should be
sufficient to determine their dimensionality and therefore the dimension of
the invariant varieties.

It would also be interesting to have examples of the unusual behaviours
which seem possible if the invariant varieties are higher dimensional
abelian varieties.  As is pointed out in eq.~(\ref{hyper}), some dynamical
exponents could be positive with a zero algebraic entropy, giving
``simple'' dynamics with diverging trajectories.  If the differential has a
more complex structure than in~(\ref{dphin}), it should also be possible to
have a polynomial growth of the degrees, but faster than quadratic. 
This would however imply a rather long recurrence relation for the
degrees, which itself is the sign of a complex scheme for the resolution of
the singularities~\cite{BeVi99}.

\appendix

\section{Birational maps.}
In this appendix, I want to recall some properties of birational maps 
which have been described in more detail in~\cite{BeVi99} and which are
relevant to this work.  

Birational maps are not really maps.
Except in the simplest case, they are not defined everywhere~\cite{Mu95}.
They are more general correspondences, which can be one
to many at some points.  But they are nevertheless maps to a good approximation
since they define maps on the complement of an algebraic variety.

The projective space $\CP^n$ is the space of complex lines through the origin
in ${\mathbb C}^{n+1}$.  Any non-zero element of the line is a set
of homogeneous coordinates for this point of projective space.  
If one of the coordinates is set to 1, the others give the so-called affine
coordinates of a part of the projective space.

A rational map between the projective spaces $\CP^m$ and $\CP^n$ is
simply a homogeneous polynomial map from ${\mathbb C}^{m+1}$ to
${\mathbb C}^{n+1}$.  The homogeneity degree is called the degree of the
map.  The image of the line representing a point in $\CP^m$ is either
identically zero in ${\mathbb C}^{n+1}$ or a complex line by the
homogeneity of the map.  In the first case, the projective point is
singular for the map and in the second, the line defines the image in
$\CP^n$.  Going to affine variables, the map is defined by rational
functions, but this leads to spurious singularities, which are just the
consequence of saying that some hyperplane of projective space is at
infinity.

The composition of rational maps requires some care.
If we compose the polynomial maps, $\phi$ of degree $d_\phi$ and $\psi$ of
degree $d_\psi$ give a map of
degree $d_\phi d_\psi$, $\psi \circ \phi$.  But with this composition
law, only maps of degree 1 can have an inverse, since $\mathbb I$ is of 
degree 1.  However, the greatest common divisor of the coordinates
of the image can be factored out and the map is not modified at the points 
where it was defined.  It is naturally extended  
to points where this divisor vanishes giving
a reduced product $\psi\times\phi$:
\begin{equation}
	\psi \circ \phi = m(\psi,\phi) \psi\times \phi.
\end{equation}
$m(\psi,\phi)$ is called a multiplier and its degree gets subtracted to the 
degree of the composed map $\psi\times \phi$.  A birational map is a rational 
map with an inverse for this $\times$-product.  The multiplier for the product
of a map and its inverse is the equation of the variety where the map is not
bijective.  When considering iterations of a rational map, we always consider 
this reduced product, which allows for a non-trivial sequence of the degrees
of the successive iterates.

\end{document}